\documentclass[twocolumn,epjc3]{svjour3}
\usepackage{graphicx}
\usepackage{subfigure}
\usepackage{bm}

 \journalname{Eur. Phys. J. C}

\begin{document}

\title{Interferometry correlations in central p+Pb collisions}

\author{Piotr Bo\.zek \and Sebastian Bysiak}

\institute{AGH University of Science and Technology, Faculty 
of Physics and Applied Computer Science, al. Mickiewicza 30,
 30-059 Krakow, Poland}
\date{Received: date / Revised version: date}
%

\maketitle

\begin{abstract}
We present results on interferometry correlations for pions emitted 
in central p+Pb collisions at 
$\sqrt{s_{NN}}=5.02$~TeV in a 3+1 dimensional viscous hydrodynamic model with 
initial conditions from the Glauber Monte Carlo  model. 
The correlation function is calculated
as a function of the pion pair rapidity. The extracted interferometry radii
show a weak rapidity dependence, reflecting the lack of boost 
invariance of the pion distribution.  
A cross-term between the $out$ and $long$ directions is found to be nonzero.
The results obtained in the hydrodynamic model are in fair agreement with recent data of the ATLAS Collaboration.
\keywords{ relativistic heavy-ion collisions \and femtoscopy \and  hydrodynamic model}

\end{abstract}

\section{Introduction}

In relativistic p+Pb collisions at the CERN Large Hadron Collider
 (LHC) a small region of
large density matter is formed, 
which makes possible the generation  of a collective flow in the 
expansion \cite{Bozek:2011if}. A number of signatures of collectivity 
due to  final state interactions have been observed experimentally 
in small collision
 systems at relativistic energies (see e.g. \cite{Loizides:2016tew}).
We use the viscous hydrodynamic model to describe the dynamics 
in p+Pb interactions. We note that initial state effect can also explain
 the observed two-particle correlations \cite{Dusling:2013qoz}.

Quantum interferometry correlations for identical particles
can serve as a measure of space-time correlations in the source
\cite{Lisa:2005dd,Wiedemann:1999qn}. These interferometry (also named Hanburry Brown-Twiss (HBT))
 correlations  have
 been studied in both elementary and  nuclear collisions. An estimate of the 
 size of 
the emission region, in the form of the HBT radii,  can be extracted from a fit
 to the interferometry correlations.
HBT radii have been measured in p+Pb collisions \cite{Adam:2015pya,CMS:2014mla} 
and can be reproduced fairly well in hydrodynamic models with Glauber model
 or color glass condensate initial conditions
 \cite{Bozek:2013df,Romatschke:2015gxa,Mantysaari:2017cni}. 

In collisions of symmetric systems the correlation function in relative 
momentum  of the pair  is usually parametrized using three HBT radii 
 \cite{Bertsch:1989vn,Pratt:1986cc}. For a source without forward-backward 
symmetry, e.g. for pairs at forward/backward rapidity, an additional cross-term 
can appear in the correlations function \cite{Chapman:1994yv}. Such a term 
is predicted to be significant in the case of asymmetric collisions
 d+Au or p+Pb 
\cite{Bozek:2014fxa}.

 Recently the ATLAS Collaboration has presented results 
on the interferometry correlations in p+Pb collisions for different 
rapidities of the pion pair 
\cite{Aaboud:2017xpw}. The HBT radii show a rapidity dependence. The
 size of the emission region is larger on the Pb going side. The cross-term 
is also  found to be nonzero, reflecting the rapidity dependence of the 
charged particle density. In this paper we present results for the
interferometry radii in central p+Pb collisions. We use the same exponential
 form of 
the correlation function and the same kinematic cuts as those used by the ATLAS
 Collaboration. We show that  the hydrodynamic model with Glauber 
model initial conditions can  semi-quantitatively  
reproduce the experimental observations.

\section{Hydrodynamic model and HBT correlations}

We describe the evolution of the matter created in p+Pb collisions using
the viscous hydrodynamic model, with initial entropy density given by 
the nucleon Glauber Monte Carlo model.
The details and the parameters of the model can be found in
 \cite{Bozek:2013df}.
The model describes well the spectra and the azimuthal flow coefficients.
At the end of the hydrodynamic evolution pions  
are emitted from the freeze-out hypersurface. 
 In the paper we
 use the rapidity in the
 nucleon-nucleon center of mass frame, which 
is shifted by $0.465$ rapidity units from the laboratory frame at the LHC.

Pairs of same-charged pions 
emitted at positions $x_1$ and $x_2$, with momenta ${\bf p_1}$ and ${\bf p_2}$,  are counted.
The three dimensional correlation function is 
constructed by binning in relative
(${\bf q}={\bf p_1}-{\bf p_2}$) and average transverse 
($k_T=|{\bf p_1}+{\bf p_2}|/2$)
  momentum of the pair, in the longitudinal comoving  system  \cite{Chojnacki:2011hb}. To increase the statistics for the pion pairs, for each 
hydrodynamic freeze-out hypersurface $500$ realistic events are 
generated (using a statistical emission procedure \cite{Chojnacki:2011hb})
 and combined together. 

We use a  symmetrized
 plane wave two-pion wave function in the definition of 
 the correlation function. Therefore, no corrections are made
 for Coulomb interaction in simulated correlation function. 
The fitted form of the correlation function is 
\begin{equation}
C({\bf q})= 1+\lambda e^{- ||R {\bf q} ||} \ ,
\label{eq:Cfit}
\end{equation}
where
\begin{eqnarray} 
|| R {\bf q}||& =& 
  \left[ (R_{out} q_{out}+R_{ol} q_{long})^2 + \right. \nonumber \\ 
& & \left. R_{side}^2 q_{side}^2+
(R_{long} q_{long}+R_{ol}q_{out})^2 \right]^{1/2}\ ,
\end{eqnarray} 
with the three components of the relative momentum $q_{out}$, $q_{side}$, and $q_{long}$  \cite{Bertsch:1989vn,Pratt:1986cc}.
The exponential ansatz (\ref{eq:Cfit}) is the same as the one  used
 by the ATLAS Collaboration \cite{Aaboud:2017xpw}, the fitted HBT radii 
$R_{out}$, $R_{size}$, and $R_{long}$ and the cross-term $R_{ol}$ are different 
than for the Gaussian ansatz used in ref. \cite{Bozek:2014fxa}.

 \begin{figure}
  \includegraphics[width=0.5\textwidth]{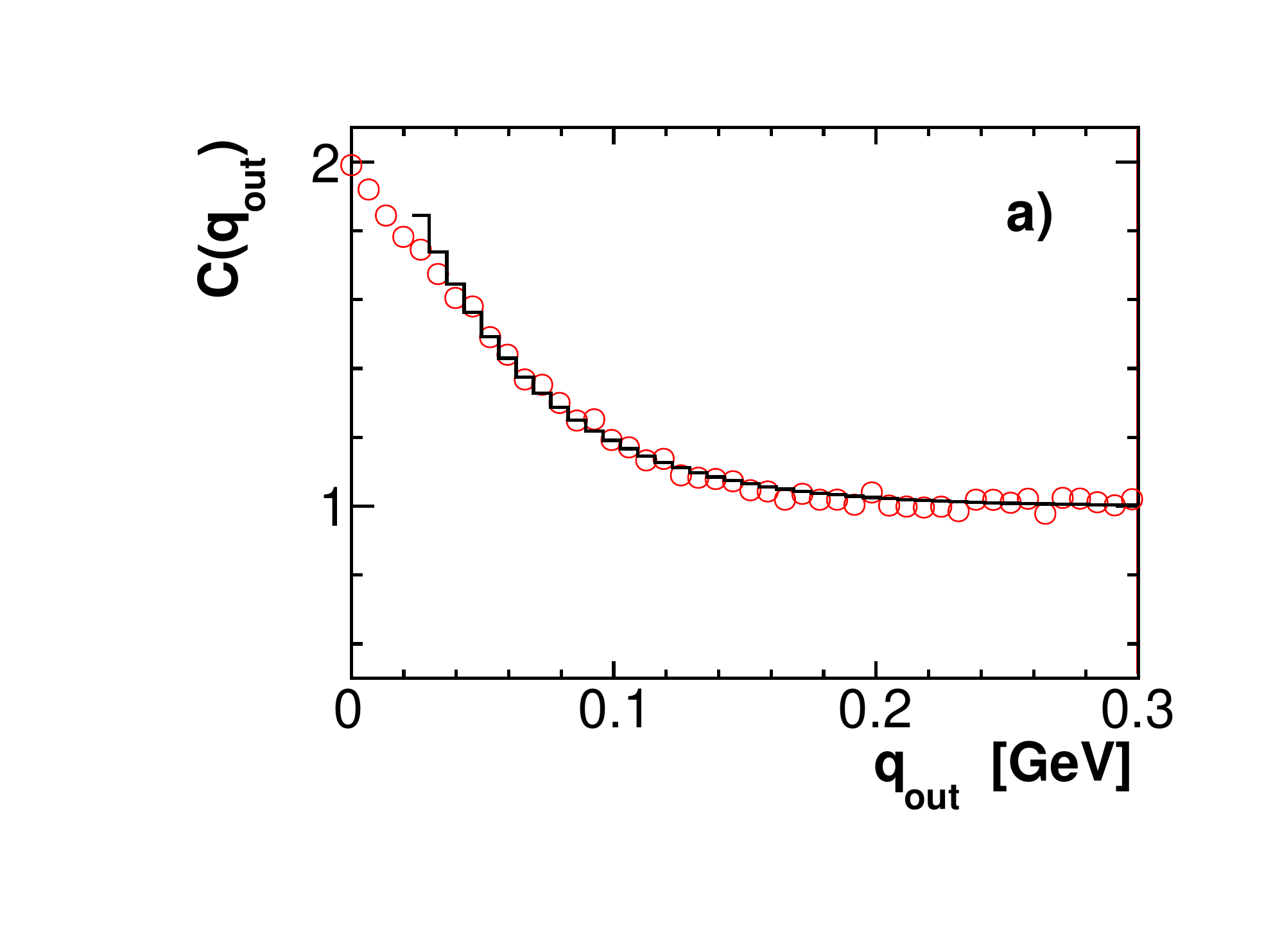}
\vskip -8mm
  \includegraphics[width=0.5\textwidth]{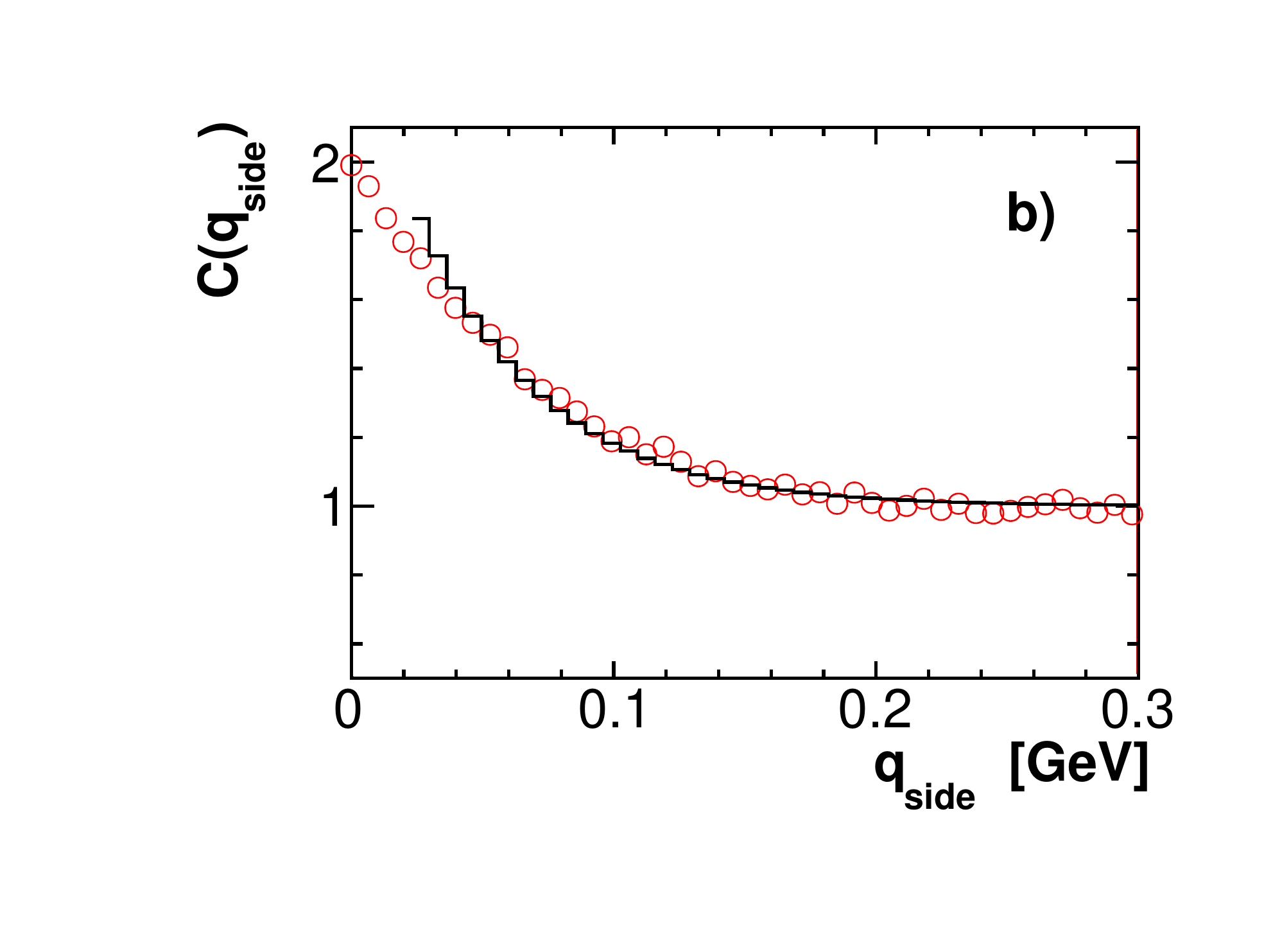}
\vskip -8mm
  \includegraphics[width=0.5\textwidth]{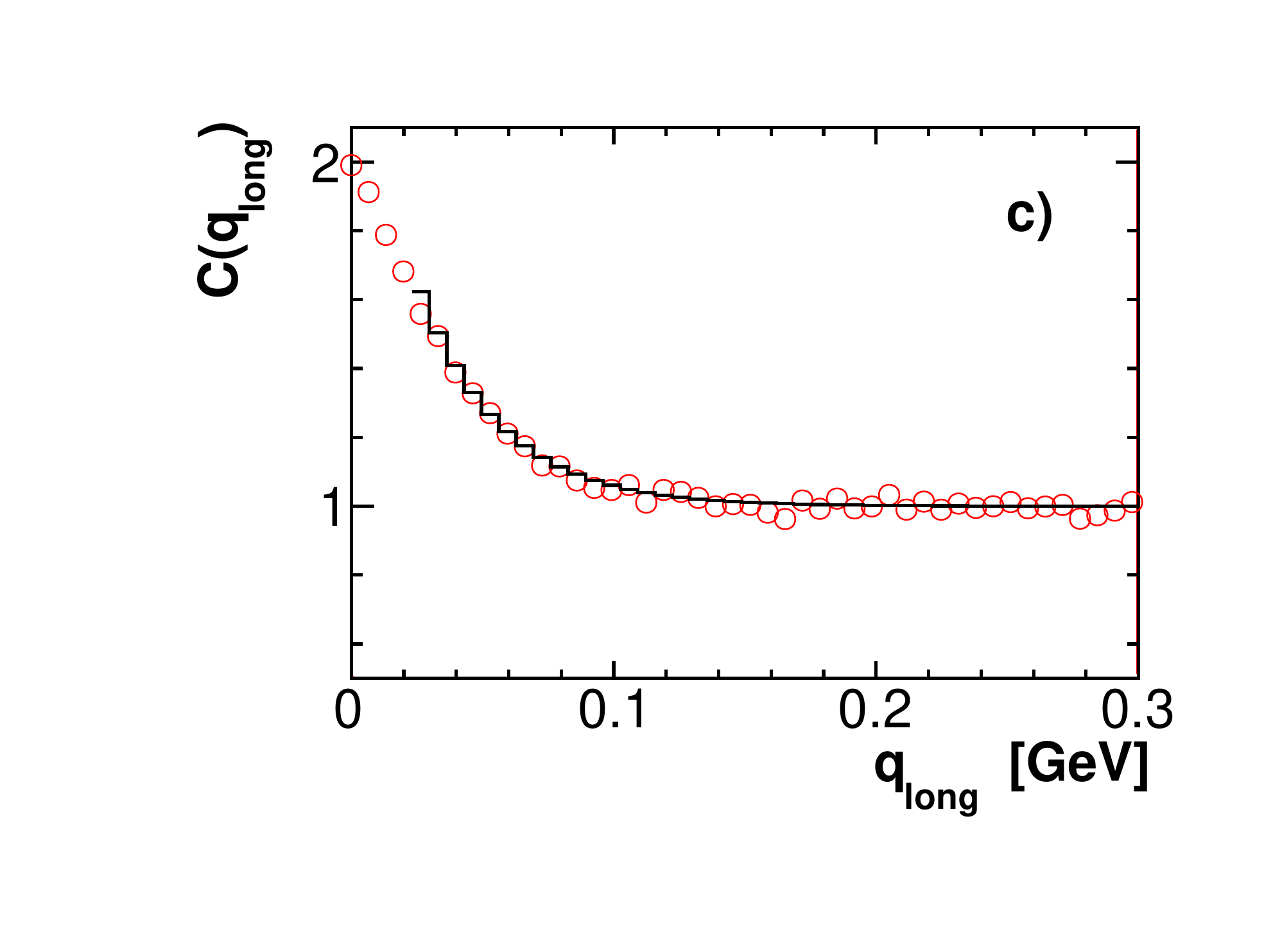}
\vskip -6mm
 \caption{Points represent the interferometry correlations for
  pions in central (0-1\%) p+Pb 
collisions, calculated in the 
viscous hydrodynamic model; $C(q_{out},q_{side}=0,q_{long}=0)$ (panel a),
 $C(q_{out}=0,q_{side},q_{long}=0)$ (panel b), and 
 $C(q_{out}=0,q_{side}=0,q_{long})$ (panel c). The solid lines represent the fitted correlation function in the region $0.025$~GeV~$<|{\bf q}|<$~$0.3$~GeV.
 }
  \label{fig:corr}
 \end{figure}

 Three cross sections of the 
interferometry correlation function are shown in Fig. \ref{fig:corr}, 
obtained in the hydrodynamic model
for the  pair transverse momentum $0.2$~GeV~$<k_T<$~$0.3$~GeV 
and rapidity $-1<y_{\pi\pi}<0$. Note that in the general case there is
 no reflection symmetry $q_{out} \leftrightarrow -q_{out}$ or 
$q_{long} \leftrightarrow -q_{long}$. Both negative and positive values 
for the ${\bf q}$ components  are taken into account for the  fits.
The correlation function is calculated for several bins 
in the average transverse momentum and rapidity of the pair $y_{\pi\pi}$. 

\section{Transverse momentum  dependent interferometry correlations}

  \begin{figure*}
  \includegraphics[width=0.5\textwidth]{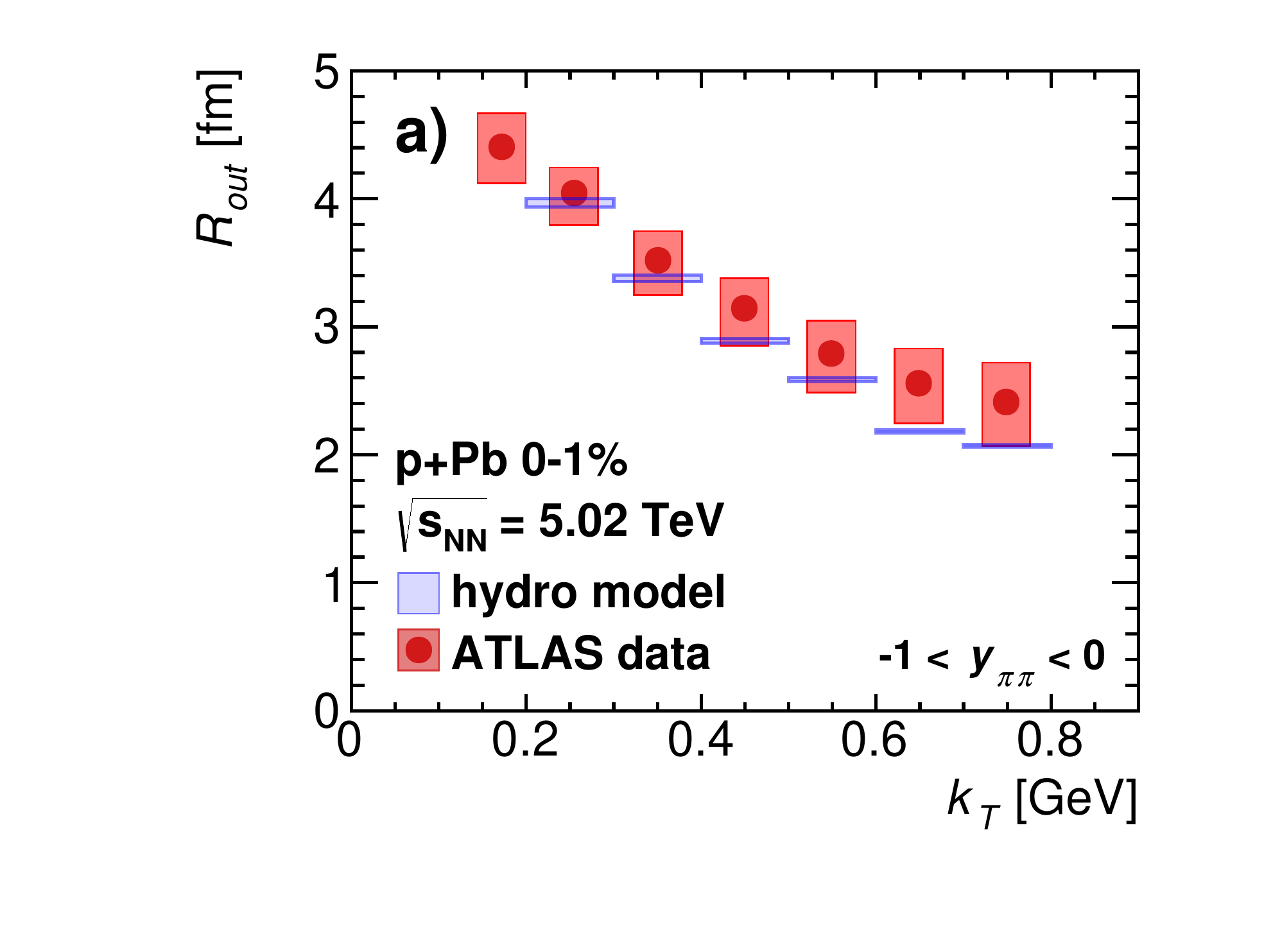}~~~\includegraphics[width=0.5\textwidth]{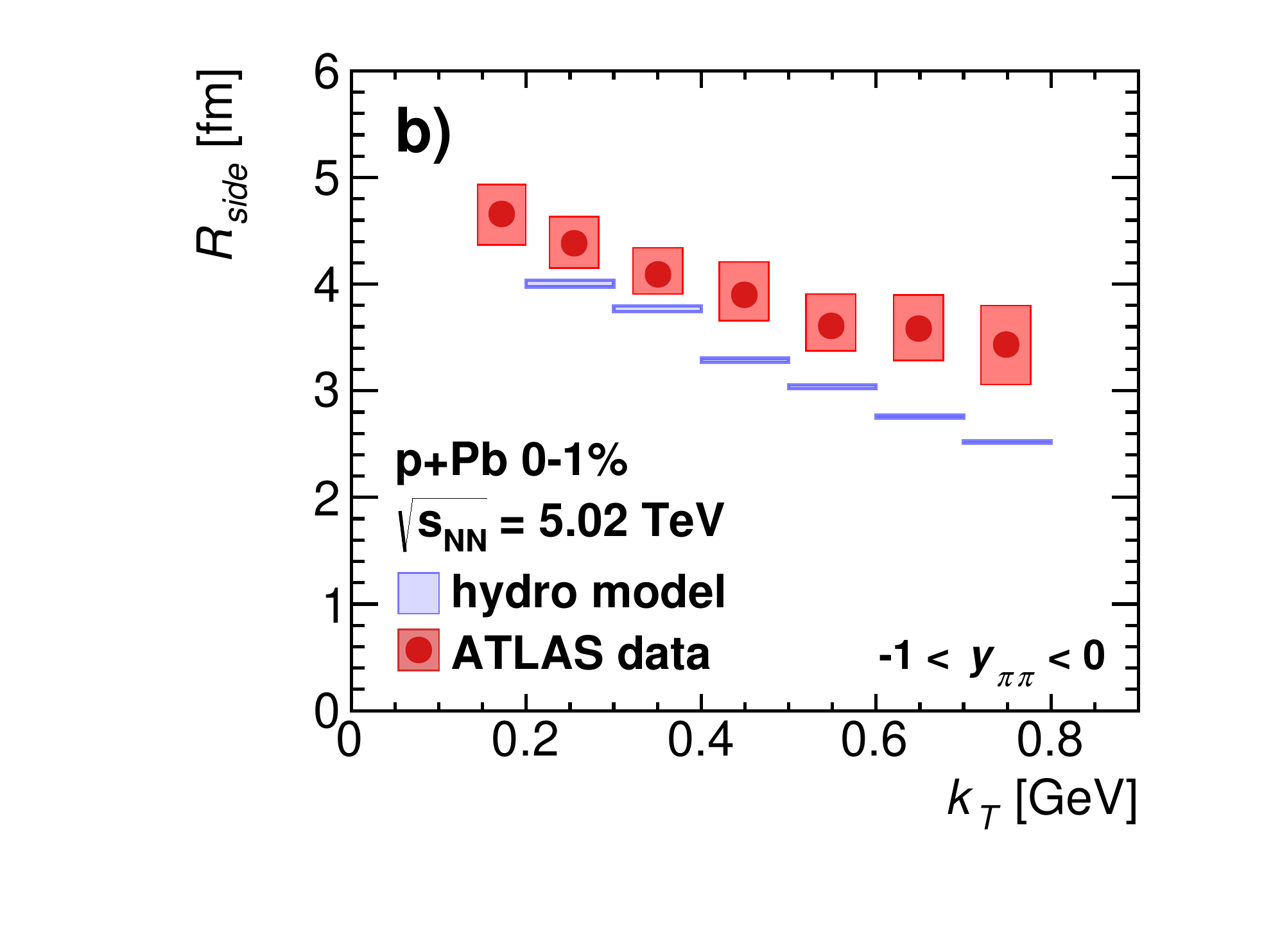}  \\
\vskip -8mm
 \includegraphics[width=0.5\textwidth]{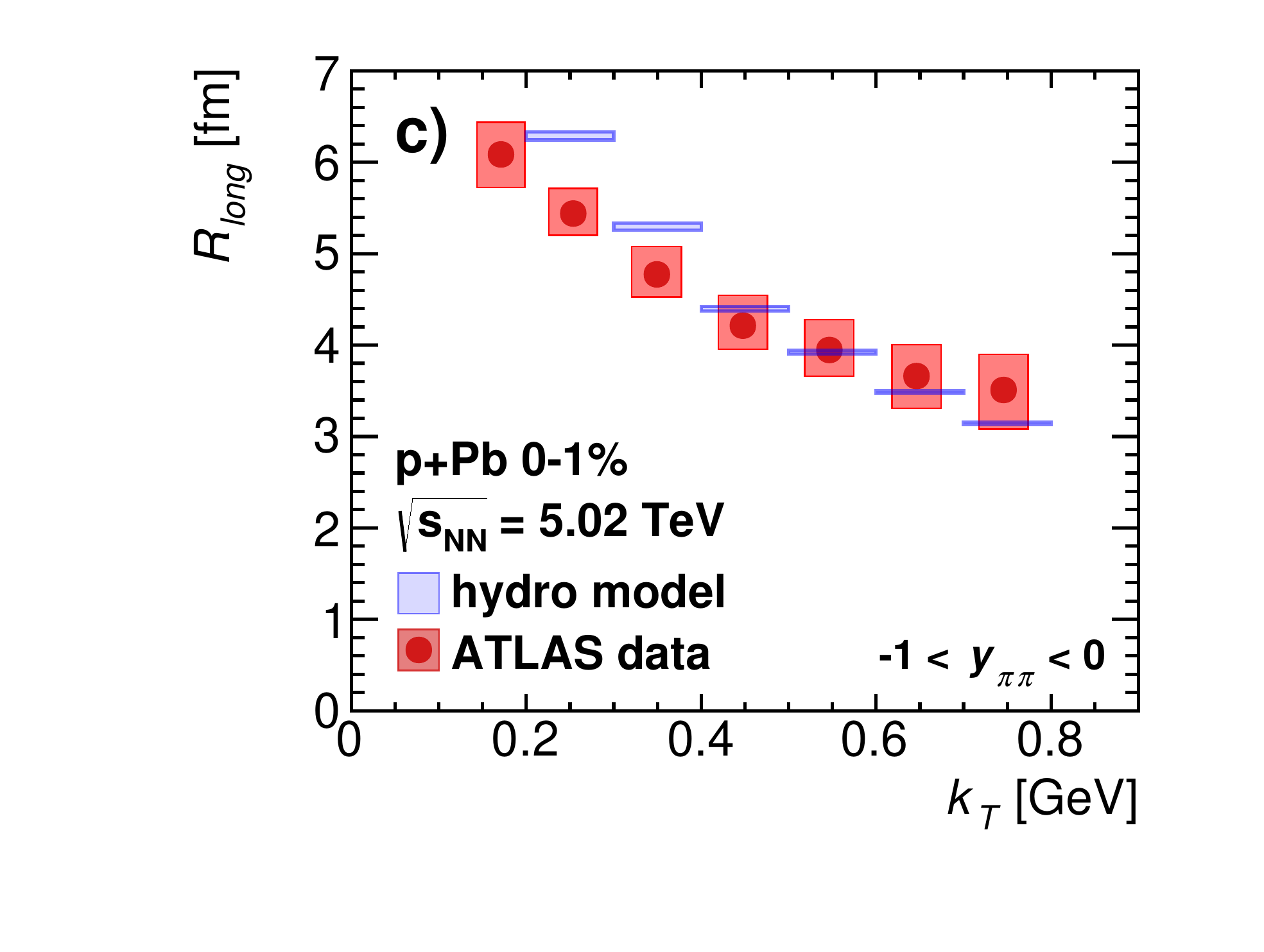}~~~\includegraphics[width=0.5\textwidth]{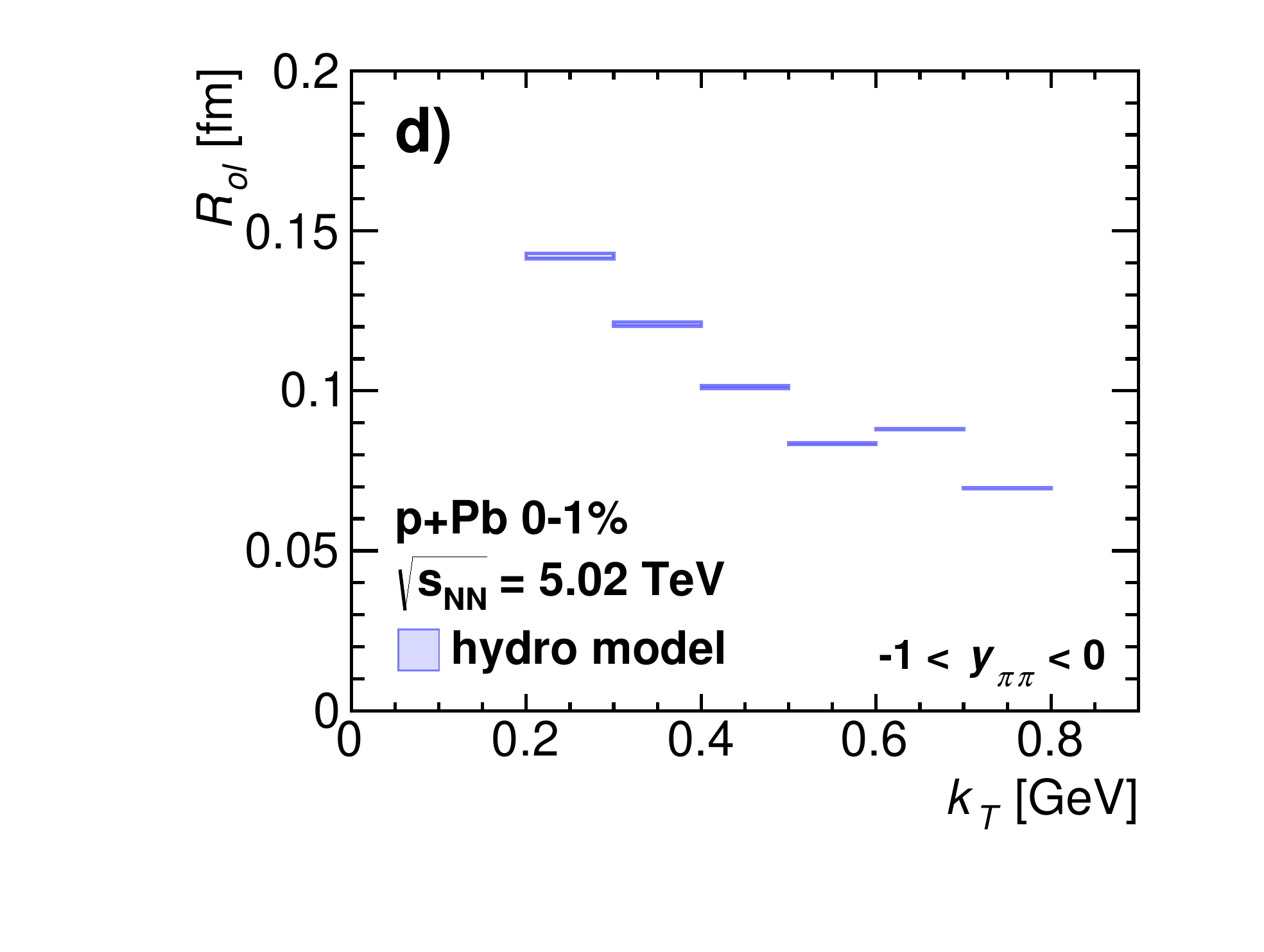}
\vskip -6mm
 \caption{Interferometry radii in central p+Pb collisions as a function of 
average pair momentum $k_T$. Results of 3-dimensional hydrodynamic calculations
are  compared to data from the ATLAS Collaboration \cite{Aaboud:2017xpw}. $R_{\rm out}$, $R_{\rm side}$, $R_{\rm long}$, and  $R_{\rm ol}$ are shown 
 in panels a) through d).} 
  \label{fig:rkt}
 \end{figure*}

In Fig. \ref{fig:rkt} are shown the three HBT radii and the cross-term $R_{ol}$ as functions
of the average transverse momentum $k_T$, for   $-1<y_{\pi\pi}<0$. 
The values of the fitted parameters depend slightly 
on the fit range for ${\bf q}$.
We have varied the range of the relative momentum in the fit 
$q_{low}<|{\bf q}|<0.3$~GeV, $q_{low}\in [0.02,0.03]$~GeV.  
The vertical size of the boxes in 
Figs. \ref{fig:rkt} and \ref{fig:rycm} represents an estimate
 of the uncertainty in the fitted parameters obtained by varying the fit range, 
for the experimental data the vertical size of the boxes represents 
the experimental
error.

The size of 
the effective emission region decreases with $k_T$. This indicates 
the existence of strong correlations  between the momentum and the
 emission point for the emitted particles.
Due to the collective flow pion pairs of high transverse momentum
 are effectively emitted from
 a smaller region of the source \cite{Akkelin:1995gh}.
The model describes well the experimental data for $R_{out}$,
 while $R_{side}$ is slightly underestimated. The pion distribution in
rapidity is not flat, which makes
 the correlations asymmetric
 in the forward backward direction. We find a nonzero cross-term $R_{ol}$, the magnitude of this term decrease with $k_T$ in a similar way as 
$R_{out}$ and $R_{long}$. The asymmetry of the interferometry correlation 
function can be related to the forward-backward asymmetry 
of the average emission times of
 pions from the source (Fig. 4 in \cite{Bozek:2014fxa}).
 Pions are emitted earlier on the proton going side and
the  correlation coefficient 
 between the  longitudinal position  and time becomes nonzero.

\section{Rapidity dependent interferometry correlations}

 \begin{figure*}
  \includegraphics[width=0.5\textwidth]{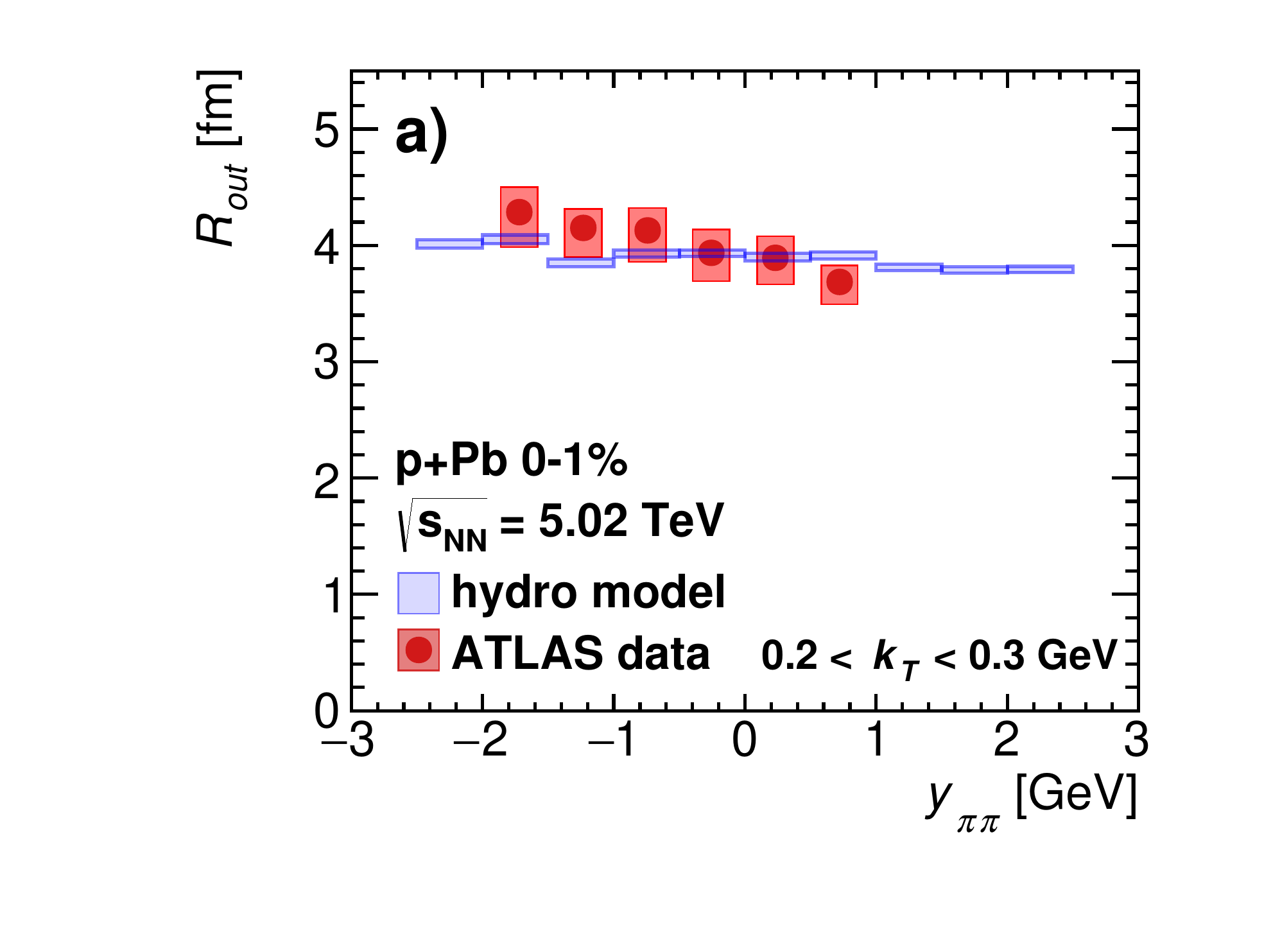}~~~\includegraphics[width=0.5\textwidth]{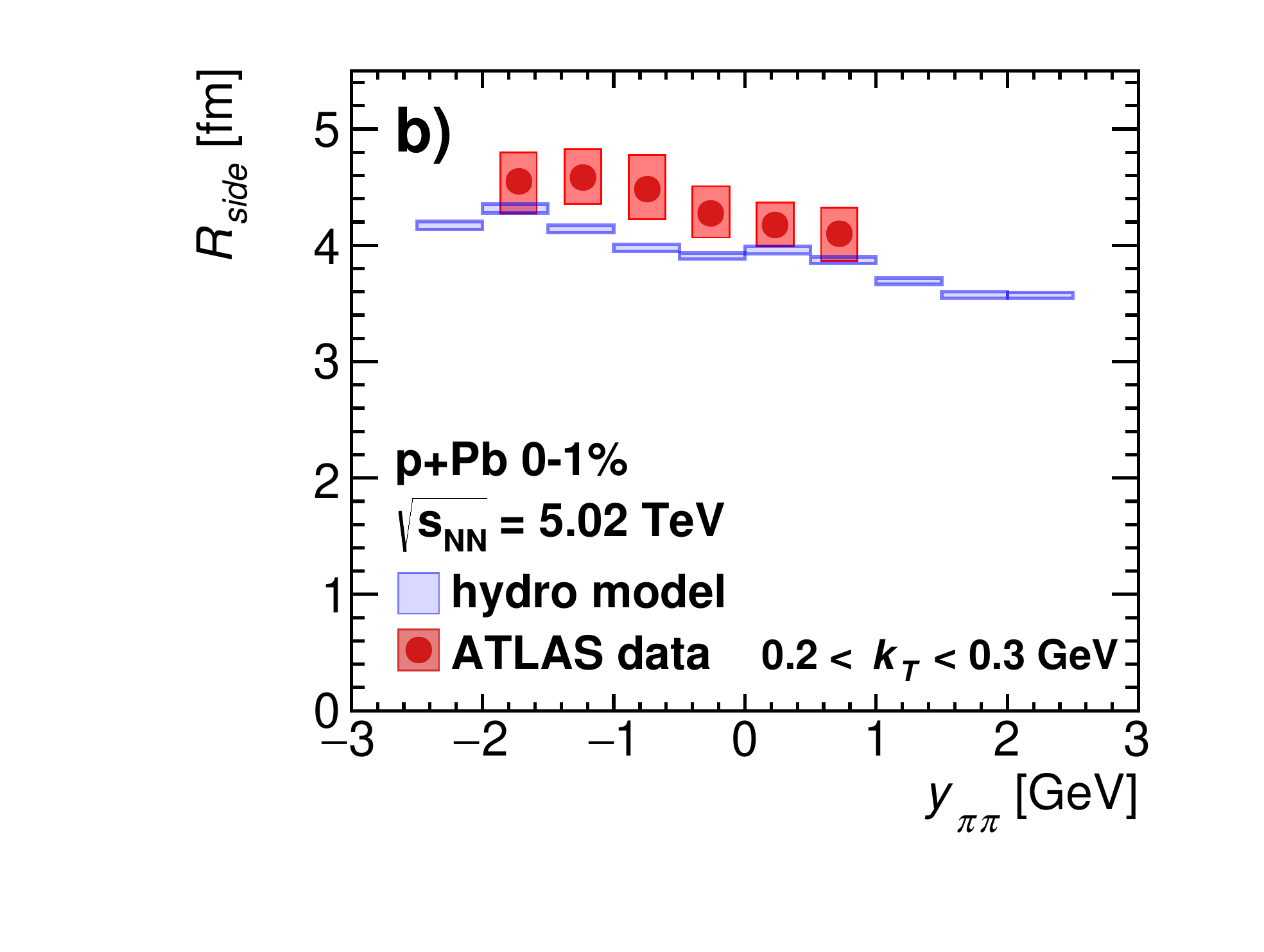}  \\
\vskip -8mm
 \includegraphics[width=0.5\textwidth]{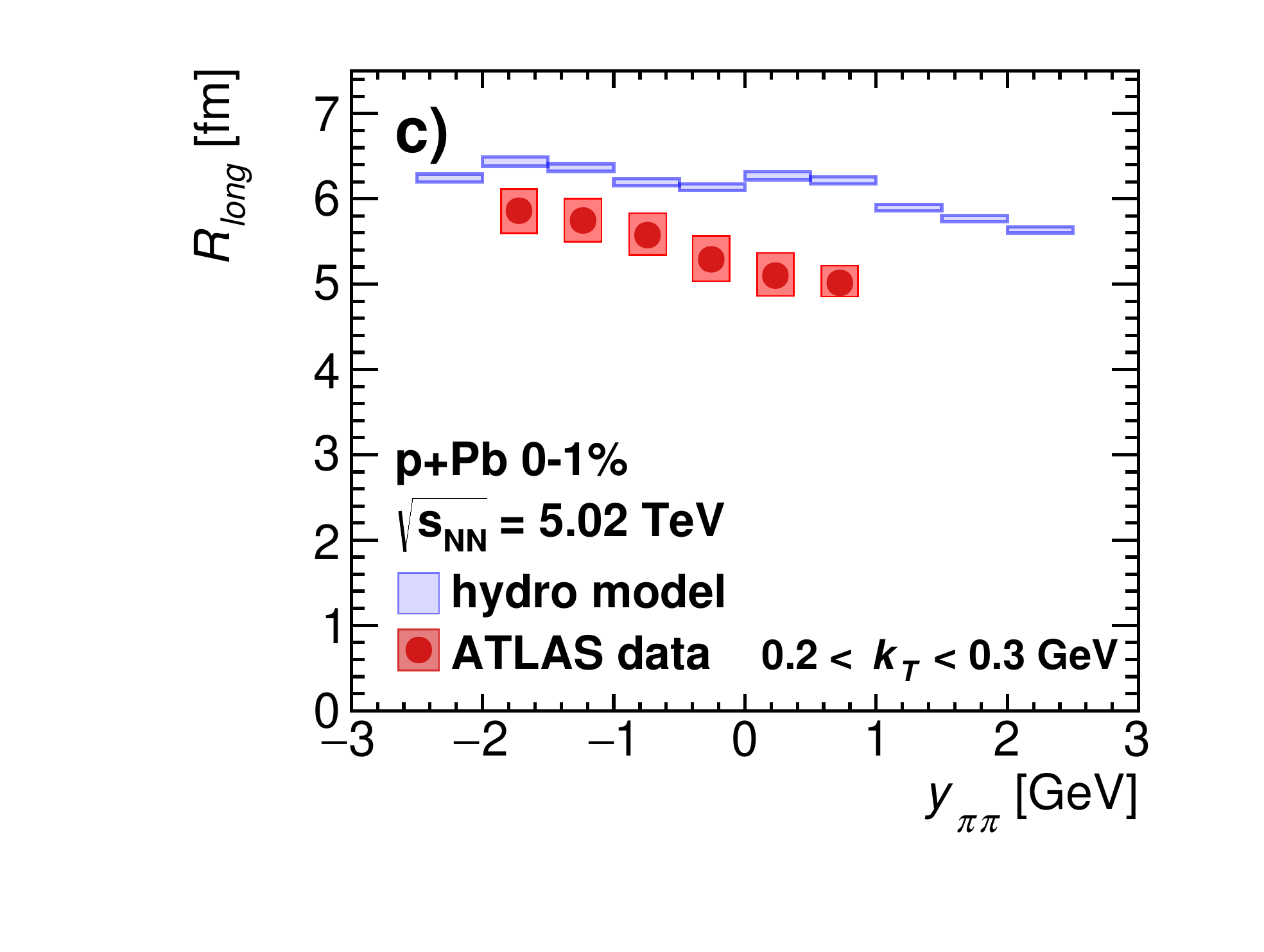}~~~\includegraphics[width=0.5\textwidth]{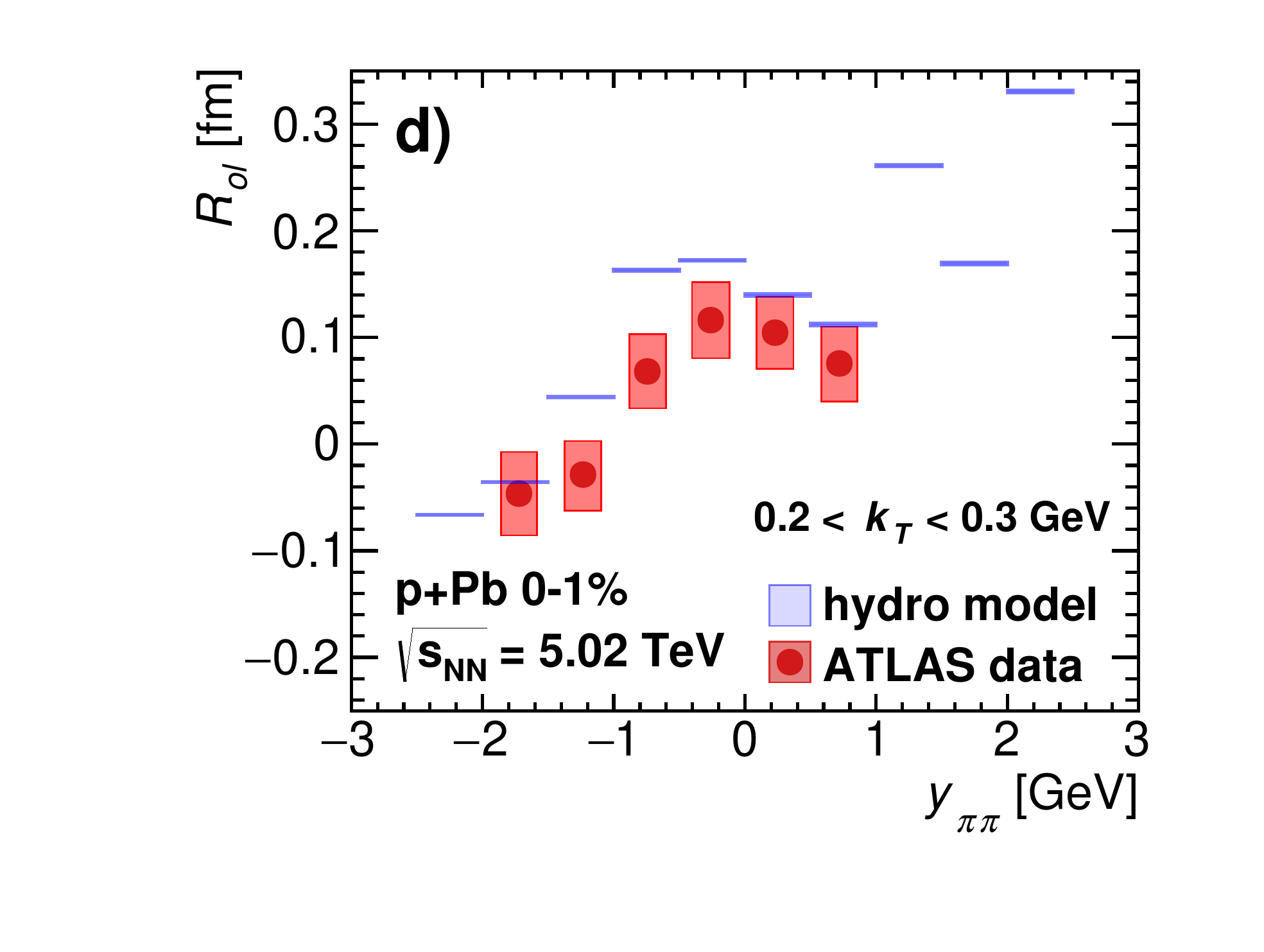}
\vskip -6mm
 \caption{Interferometry radii in central p+Pb collisions as a function of 
average pair rapidity $y_{\pi \pi}$. Results of 3-dimensional hydrodynamic calculations are compared to data from the ATLAS Collaboration \cite{Aaboud:2017xpw}. $R_{\rm out}$, $R_{\rm side}$, $R_{\rm long}$, and  $R_{\rm ol}$ are shown in panels a) through d).} 
  \label{fig:rycm}
 \end{figure*}

In Fig. \ref{fig:rycm} presents  the rapidity dependence of the HBT parameters and  the $R_{ol}$ cross-term for pion pairs with $0.2$~GeV~$<k_T<0.3$~GeV.
Both the experimental data and the simulation results show some dependence 
on the
pion pair rapidity $y_{\pi\pi}$. However, the experimentally 
observed rapidity dependence is slightly stronger. We notice that the model
 correctly  describes the experimental data for the cross-term $R_{ol}$
  (panel d) in Fig. \ref{fig:rycm}). Both the magnitude and the sign of the 
cross-term are fairly well reproduced.

\begin{figure}
  \includegraphics[width=0.5\textwidth]{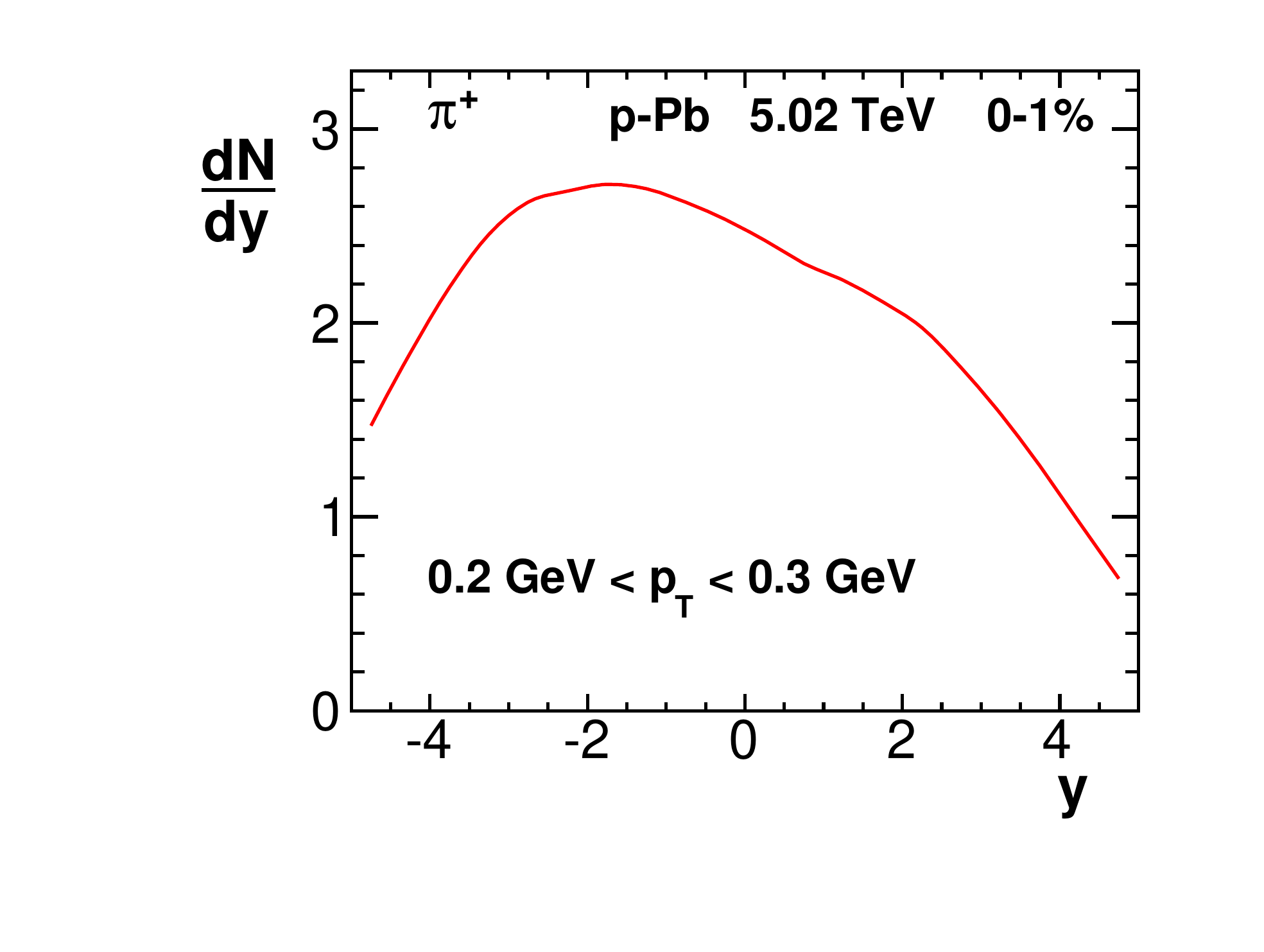}
\vskip -6mm
 \caption{Rapidity distribution of $\pi^+$ in central (0-1\%) p+Pb 
collisions with $0.2$~GeV$<p_T<0.3$~GeV, calculated in the 
viscous hydrodynamic model with Glauber Monte Carlo model initial conditions.
 }
  \label{fig:pirap}
 \end{figure}

The rapidity distribution of pions emitted in central p+Pb collisions
 is far from boost invariant (Fig. \ref{fig:pirap}). 
The asymmetry results from 
the large difference in the number of participants between the two projectiles,
 one for the proton and $19$ on average for the Pb nucleus. 
The change in the number of emitted pions with rapidity determines the change in the effective size of the emission region as measured through the HBT radii.
 The sign  of the cross-term $R_{ol}$ 
follows
approximately the reverse sign of the slope of the rapidity distribution of pions.

\section{Summary}

We analyze the interferometry correlations for pions emitted in central p+Pb 
collisions at the LHC. The correlations functions are calculated
 in the viscous hydrodynamic model with Glauber Monte Carlo initial conditions.
The correlation function is constructed from same-charged pion pairs. Three 
interferometry radii are extracted from an exponential fit to the 
correlation function. The lack of forward-backward symmetry in the emission
 region leads to an additional cross-term coupling the $q_{out}$ 
and $q_{long}$ directions. The corresponding cross-term parameter $R_{ol}$ is
 found to be nonzero. 

The interferometry radii $R_{out}$, $R_{side}$, $R_{long}$, and the cross-term
$R_{ol}$ are calculated as functions of the average transverse momentum $k_T$ and rapidity $y_{\pi\pi}$ of the pion pair. The model reproduces the recent ATLAS
 Collaboration data to $10$\% accuracy for the $k_T$ dependence 
of the HBT parameters.
We find that the HBT radii as functions of the
 pair rapidity $y_{\pi\pi}$ are smaller on proton going side. The rapidity
dependence of the cross-term follows  the slope of the rapidity
 distribution of emitted pions and is in good agreement with experimental data.
The calculation confirms that the hydrodynamic model can reproduce qualitatively
the space-time  features of the emission source produced in high energy 
p+Pb collisions.


\begin{acknowledgements}
Supported in part by Polish Ministry of Science and Higher Education (MNiSW),
 by 
National Science Centre, grant  2015/17/B/ST2/00101, and  by PL-Grid Infrastructure.  
\end{acknowledgements}

\bibliography{../hydr}

\end{document}